\documentstyle[sprocl]{article}
\input epsf

\input{psfig}

\def\Journal#1#2#3#4{{#1} {\bf #2}, #3 (#4)}

\def\NPB{{\em Nucl. Phys.} B}
\def\NPA{{\em Nucl. Phys.} A}
\def\PLB{{\em Phys. Lett.}  B}
\def\PRL{\em Phys. Rev. Lett.}
\def\PRD{{\em Phys. Rev.} D}
\def\PRC{{\em Phys. Rev.} C}

\def\half{{1\over 2}}
\def\MeV{{\rm MeV}}

\def\Tr{{\rm Tr\,}}
\def\nrcpt{NR\raise.4ex\hbox{$\chi$}PT\ }

\def\ltap{\ \raise.3ex\hbox{$<$\kern-.75em\lower1ex\hbox{$\sim$}}\ }
\def\gtap{\ \raise.3ex\hbox{$>$\kern-.75em\lower1ex\hbox{$\sim$}}\ }

\def\CA{{\cal A}}

\def\CL{{\cal L}}

\def\pds{{\it PDS}\ }

\def\bfq{{\bf q}}

\def\frac#1#2{{\textstyle{#1\over#2}}}
\def\darr#1{\raise1.5ex\hbox{$\leftrightarrow$}\mkern-16.5mu #1}
\def\){\right)}
\def\({\left(}
\def\]{\right]}
\def\[{\left[}
\def\si{{}^1\kern-.14em S_0}
\def\siii{{}^3\kern-.14em S_1}
\def\diii{{}^3\kern-.14em D_1}
\def\fm{{\rm\ fm}}
\def\MeV{{\rm\ MeV}}
\def\CA{{\cal A}}

\def\be{\begin{equation}}
\def\ee{\end{equation}}
\def\bea{\begin{eqnarray}}
\def\eea{\end{eqnarray}}
\begin{document}

\title{WHAT EFFECTIVE FIELD THEORY MAY CONTRIBUTE TO THE BLAST PROGRAM\footnote{Talk presented at the
    {\it Second Workshop on Electronuclear Physics with Internal Targets
      and the Bates Large Acceptance Spectrometer Toroid (BLAST)},  May 1998.}
\footnote{NT@UW-18}
}

\author{{Martin  J. Savage
}}

\address{
Department of Physics, University of Washington,  Seattle, WA 98195}

\maketitle
\abstracts{
Recent progress in using effective field theory
to describe two nucleon systems is reviewed.
}

%%%%%%%%%%%%%%%%%%%%%%%%%%%%%%%%%%%%%%%%%%
\section{Introduction}
\label{sec:1}
%%%%%%%%%%%%%%%%%%%%%%%%%%%%%%%%%%%%%%%%%%

Several years of effort\cite{Weinberg1}$^-$\cite{KSWnew}
has culminated in a consistent effective field
theory description of the nucleon nucleon interaction\cite{KSWb}.
The ultimate goal of this endevour is to construct a framework
with which to systematically describe bound and unbound
multi-nucleon systems as well as elastic and inelastic processes.
This effort was initiated by Weinberg's pioneering work on the 
subject~\cite{Weinberg1}\ where he proposed a power-counting
scheme for local-operators involving two or more nucleons
and the inclusion of pions.
However, it was shown that Weinberg's power counting scheme is not consistent.
Recently, a consistent 
power-counting scheme has been proposed\cite{KSWb} to describe
$NN$ scattering and applied to observables of the deuteron.
I will focus on these subjects in this talk and 
give an indication of what this program of study might acheive
in the near future.

%%%%%%%%%%%%%%%%%%%%%%%%%%%%%%%%%%%%%%%%%%
\section{Why Effective Field Theory?}
\label{sec:2}
%%%%%%%%%%%%%%%%%%%%%%%%%%%%%%%%%%%%%%%%%%

Effective field theory is a very powerful technique for dealing with systems
that possess widely seperated length scales.
In a system with just two length scales $l_1$ and $l_2$ (as an example),
the ratio ${\cal Q}=l_1/l_2$ can be used as a small expansion parameter.
Usually systems possess more symmetries when ${\cal Q}=0$ and for
small $Q$ a perturbative expansion exists that makes use of the
symmetries of the unperturbed system.

In the theory of strong interactions there is one intrinsic length scale,
$\Lambda_{\rm QCD}$.
For processes that occur at distance scales that are small compared to 
$\Lambda_{\rm QCD}$ the appropriate
degrees of freedom are the quarks and gluons with
the QCD Lagrange density used to compute processes
as a perturbative expansion in the strong
interaction coupling constant, $\alpha_s(\mu)$.
In addition, there is power series expansion in
forming the  matrix elements of the quark-gluon operators, with an expansion parameter
$\Lambda_{\rm QCD}/Q$.
It is often useful to impose the constraints of chiral symmetry, arising from
the smallness of the light quark masses compared to $\Lambda_{\rm QCD}$,
giving yet a third expansion parameter, $m_q/\Lambda_{\rm QCD}$.
An explicit example is the semileptonic decay of
b-flavored hadrons. The inclusive decay rate for the decay of a $B$-meson to a charmed
final state is\cite{FLSa}
\begin{eqnarray}
  \Gamma (B\rightarrow X_cl\overline{\nu}) & = &
  {G_F^2 M_B^5 |V_{cb}|^2\over 192\pi^3}  0.369
  \left[  1 - 1.54 {\alpha (m_B)\over \pi}
    - 1.43 \left( {\alpha (m_B)\over \pi} \right)^2\beta_0
    \right.
    \nonumber\\
    & & \left.
    - 1.65 {\overline{\Lambda}\over M_B} \left( 1-0.87 {\alpha (m_B)\over
        \pi}\right)
    \right.
    \nonumber\\
    & & \left.
    -0.95  \left({\overline{\Lambda}\over M_B}\right)^2
    - 3.18 {\lambda_1\over M_B^2} + 0.02 {\lambda_2\over M_B^2}
    \right]
\ \ \ \ .
\end{eqnarray}
The quantities $\overline{\Lambda}$, $\lambda_1$ and $\lambda_2$ are
nonperturbative matrix elements corresponding to the 
energy of the light degrees of freedom of the $B$-meson,
the $b$-quark fermi-motion and the chromomagnetic
interaction between the  $b$-quark and the  light degrees of freedom,
respectively.
In contrast, for processes at intermediate length scales there are no small expansion
parameters.
The strong interaction coupling $\alpha_s (\mu)$ is approaching unity, as is
$\Lambda_{\rm QCD}/Q$.
At long distances the appropriate degrees of freedom
are the hadrons themselves,
and not the quark and gluon fields.
$U(1)_{\rm em}$ gauge symmetry of
electromagnetism, the $SU(2)_L\otimes SU(2)_R$ chiral symmetry
and the small expansions parameters
${\bf p}/\Lambda$ and $m_q/\Lambda$ (where ${\bf p}$ denote the external
momentum of the hadrons, $m_q$ are the light quark masses and $\Lambda$ is the
scale of strong interactions)
can be encorporated  a
Lagrange density describing the low energy dynamics of hadrons and photons.

Writing the  Lagrange density as
\begin{eqnarray}
  {\cal L}\left( {{\bf p}\over\Lambda},{m_q\over \Lambda}\right)
& = & \sum_i\ C^{(i)} (\mu)\ {\cal O}^{(i)} (\mu)
\ \ \ ,
\label{eq:sep}
\end{eqnarray}
where the operators ${\cal O}^{(i)} (\mu)$ are renormalized at the scale $\mu$
and are constructed from the hadronic fields.
Coefficients $ C^{(i)} (\mu)$ renormalized at
$\mu$ are determined by the short-distance behavior of the strong interactions.
Clearly eq.(\ref{eq:sep}) represents an explicit  seperation of scales.
By construction, observables do not depend upon the scale $\mu$ at which one chooses
to renormalize.
Initially, this appears to be a disasterous situation, since
there are an infinite number of operators
that one can construct and hence there are an infinite number of
constants unconstrained by the symmetries of the theory.
However, we have learned much about dealing with
non-renormalizable field theories over the last many years and in general
such theories are  predictive.
To construct relations between certain observables that are valid
to a certain precision, only a finite
number of constants in the effective Lagrange density need to be determined.
The key to making such theories predictive is to establish a consistent power counting
scheme, one in which terms that ``look small'' at the level of the
lagrange density do in fact make small contributions to observables, even
with the inclusion of loop effects.

%%%%%%%%%%%%%%%%%%%%%%%%%%%%%%%%%%%%%%%%%%
\section{The $NN$ Interaction and Weinberg's Power Counting}
\label{sec:3}
%%%%%%%%%%%%%%%%%%%%%%%%%%%%%%%%%%%%%%%%%%

The most general Lagrange density consistent with chiral symmetry describing the
interaction of two nucleons is 
\bea
 \CL & = &  N^\dagger (iD_0+ \vec D^2/2M) N 
 + {f^2\over 8} \Tr \partial_\mu\Sigma^\dagger  \partial^\mu \Sigma +
{f^2\over 4}\omega\Tr m_q (\Sigma + \Sigma^\dagger)  
\nonumber\\
& -& \half C_S (N^\dagger N)^2 
-\half C_T (N^\dagger \vec\sigma N)^2 
+ g_A N^\dagger \vec A\cdot\vec\sigma N
\ \ +\ \ ...
\ \ \ ,
\label{eq:weinL}
\eea
where $D_\mu$ denotes a chiral-covariant derivative and
$\Sigma$ is the exponential of the isotriplet of pions
\bea
\Sigma & = & \exp \left({2i\over f} M\right)
\nonumber\\
M & = & \left( 
\matrix{\pi^0/\sqrt{2} & \pi^+\cr  \pi^- & -\pi^0/\sqrt{2} }
\right)
\ \ \ ,
\eea
with $f=132\ {\rm MeV}$ the pion decay constant and 
$\vec A$ is the axialvector meson field.
The ellipses denote
terms with more spatial derivatives and more insertions of the 
light quark mass matrix, $m_q$.
The Georgi-Manohar\cite{GeorgiManohar} 
naive dimensional analysis arising from a consideration of 
loop contributions to observables suggests that 
$C_{S,T}\sim 1/f^2$, 
and Weinberg's power-counting will follow directly.

A necessary ingredient for an EFT is a power counting scheme that dictates
which graphs to compute in order to determine an observable 
to a desired order in the expansion.
We denote the expansion parameters of the theory by $Q\sim |{\bf p}|, m_q^{1/2}$.
The main complication in the theory of nucleons and pions is
the fact that a nucleon propagator 
$S(q)=i/(q_0 - {\bf q}^2/2M)$ scales like $1/Q$ if $q_0$ scales like 
$m_\pi$ or an external 3-momentum,    
while $S(q)\sim M/Q^2$ if $q_0$ scales like an external kinetic
energy.  
Similarly, in loops $\int {\rm d} q_0$ can scale like $Q$ or $Q^2/M$,
depending on which pole is picked up. 
To distinguish between these two
scaling properties
it is convenient to  define generalized ``$n$-nucleon potentials'' $V^{(n)}$
comprised of those parts of connected Feynman diagrams with
$2n$ external nucleon lines that have no powers of $M$ in their scaling 
(except from relativistic corrections). 
$V^{(n)}$ includes  diagrams which are $n$-nucleon irreducible 
and  parts of diagrams which are 1-nucleon irreducible.
To compute the latter contribution  to $V^{(n)}$
one identifies all combinations of two or more internal nucleon lines
that can be simultaneously on-shell, and excludes their pole
contributions when performing the $\int{\rm d}q_0$ loop integrations.

Two nucleon scattering is simple since the
graphs are all ladder diagrams with insertions of  $V^{(2)}$'s acting
as ladder rungs.   
Each loop of the ladder introduces a loop integration (${\rm
d}q_0{\rm d}^3\bfq\sim Q^5/M$) and two nucleon propagators ($ M^2/Q^4$) to
give a  factor of $(QM)$ per loop.
If one treats $M\simeq Q^0$, it
follows that perturbation theory is adequate for describing
the 2-nucleon system at low $Q$. 
In order to accommodate large scattering lengths and bound states near threshold,
as in the $\si$ and $\siii-\diii$ channels
one must conclude that $ M\sim 1/Q$ in this power counting
scheme.  
At leading order one must sum up all ladder diagrams with
insertions of the leading two-body potential $V^{(2)}_0$, while  at
subleading order one includes one insertion of  the subleading potential, $V^{(2)}_1$
and all powers of $V^{(2)}_0$, and so forth.
\begin{figure}[t]
\qquad\qquad\psfig{figure=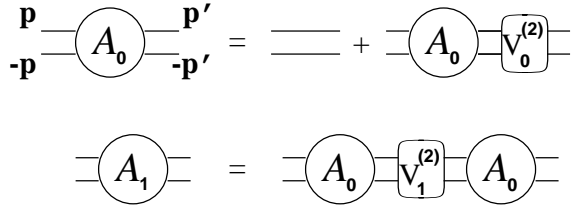,height=1.2in}
\caption{
The first two terms in the  EFT expansion of the Feynman amplitude
for nucleon-nucleon scattering in Weinberg's
power-counting scheme.  
The leading amplitude $\CA_0$ consists of the sum of ladder diagrams with the leading
2-nucleon potential $V^{(2)}_0$ at every rung.  
\label{wein_fig2}}
\end{figure}

\begin{figure}[t]
\qquad\qquad\qquad\psfig{figure=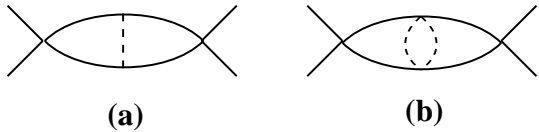,height=0.8in}
\caption{Graphs  with logarithmic divergences.
The divergence in graph (a) is proportional to $M^2 m_\pi^2$,
while graph (b) has a divergence  proportional to  $M^2 {\bf p}^4$. 
The solid lines are nucleons and the dashed lines are pions.
\label{wein_fig3}}
\end{figure}

At leading order in Weinbergs power-counting there are contributions to 
$V_0^{(2)}$ from both the local four-nucleon operators, $  C_{S,T}$ and 
from the exchange of a single potential pion,
giving a momentum space potential of 
\be
V_0^{(2)}({\bf p},{\bf p^\prime}) = C \ - \ \({g_A^2\over
2f_\pi^2}\){({\bf q} \cdot\sigma_1   {\bf q} \cdot\sigma_2)(\tau_1\cdot\tau_2)
\over ({\bf q}^2+m_\pi^2)\ }
\ \ \ \ ,
\label{eq:ope}
\ee
where $C$ denotes the  combination of $C_{S,T}$ appropriate for a
given spin-isospin channel.
The leading order amplitude results from summing the graphs 
shown in Fig.~(\ref{wein_fig2})
resulting  from this  potential, 
i.e. solving the Schrodinger equation.
In the $\si$ channel at two-loops in the ladder sum there is a  
logarithmic divergence in the graph shown in Fig.~(\ref{wein_fig3}a)
that  must be regulated.
In dimensional regularization the divergent part of this graph is 
\be
-{1\over \epsilon} {g_A^2 m_\pi^2 M^2\over 128\pi^2 f_\pi^2} C^2
\ \ \  ,
\label{eq:pole}
\ee
which requires a counterterm with a single insertion of the light 
quark mass matrix.
However, the coefficients of these operators must scale like $M^2$, and since
$ m_\pi^2 M^2 \sim Q^0$
these formally higher order operators in 
Weinberg's power-counting are required at leading order to absorb 
divergences in the time-ordered products of the leading order potential,
$V_0^{(2)}$.
Ignoring the multi-pion vertices arising from these operators, 
they can be re-absorbed into the leading operators with coefficients
$C_{S,T}$.

\begin{figure}[t]
\qquad\qquad\qquad\psfig{figure=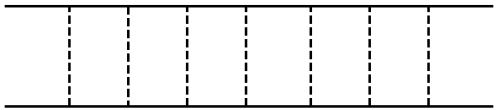,height=0.6in}
\caption{A contribution to the pion ladder sum, arising at leading 
order in Weinberg's power-counting.
The solid lines are nucleons and the dashed lines are pions.
\label{wein_fig4}}
\end{figure}

The situation is different in the $\siii-\diii$ channel and in higher partial waves.
A contribution to the leading order ladder sum is shown in Fig.~(\ref{wein_fig4}),
arising from seven potential pion exchanges, i.e a six-loop graph.
It is straightforward to deduce that
this graph has a logarithmic divergence at order $(QM)^6$, and therefore, 
counterterms
involving $\nabla^6$ 
are required at leading order in the expansion.
Clearly, the same discussion can be made for an arbitrary number of 
potential pion exchanges, and therefore counterterms involving an arbitrary even 
number of $\nabla$'s are required.
This is a clear demonstration of the failure of Weinbergs power-counting.
Further, this conclusion is true for all regularization schemes 
and not just for dimensional regularization.

%%%%%%%%%%%%%%%%%%%%%%%%%%%%%%%%%%%%%%%%%%
\section{A New Power Counting}
\label{sec:4}
%%%%%%%%%%%%%%%%%%%%%%%%%%%%%%%%%%%%%%%%%%

Lets us begin by examining the general form of the amplitude
for nucleon scattering in a S-wave
\bea
\CA & = & {4\pi\over M} {1\over p \cot\delta - i p}
\ \ \ .
\label{eq:ampa}
\eea
From quantum mechanics it is well known that 
$p\cot\delta$ has a momentum
expansion for $p\ll\Lambda$ (the effective range expansion),
\be
p\cot\delta = -{1\over a} + {1\over 2}\Lambda^2\sum_{n=0}^\infty {r}_n
\({p^2\over \Lambda^2}\)^{n+1}
\ \ \ \ ,
\label{eq:erexp}
\ee
where $a$ is the scattering length, and 
$r_0$ is the effective range.
For scattering in the $\si$ and $\siii$ channels the scattering lengths
are found to be large, $a^{(\si)} = -23.714\pm 0.013\ {\rm fm}$ and 
$a^{(\siii)} = +5.425\pm 0.0014\ {\rm fm}$ respectively.
Expanding the expression for the amplitude in eq.(\ref{eq:ampa}) 
in powers of $p/\Lambda$ while retaining $ap$ to all orders gives
\be
\CA = -{4\pi\over M}{1\over (1/a + i p)}\[ 1 + {r_0/2 \over (1/a + ip)}p^2 +  
{(r_0/2)^2\over (1/a + ip)^2} p^4 + {(r_1/2\Lambda^2)\over (1/a + ip) } p^4 +\ldots\]
\label{eq:aexp2}
\ee
For $p>1/|a|$ the terms  in this expansion scale as 
$\{p^{-1}, p^0,p^1,\ldots\}$, and 
the expansion in the effective theory takes the form
\be
\CA=\sum_{n=-1}^\infty \CA_n 
\ \ \  ,\ \ \ \CA_n \sim p^n
\ \ \ \ .
\label{eq:ampbiga}
\ee

In the theory without pions we can explicitly compute the 
s-wave amplitude in each spin channel
to all orders in the momentum expansion,
\bea
\CA & = & -{  \sum C_{2n} p^{2n} 
\over
1 + M(\mu+ip)/4\pi \sum C_{2n} p^{2n} }
\ \ \ \  ,
\label{eq:answer}
\eea
where $C_{2n}$ is the coefficient of the $p^{2n}$ term in the lagrange density.
$\mu$ is the renormalization scale and we have used 
Power Divergence Subtraction (\pds)\cite{KSWb} to define the  theory.
A typical loop graph that appears in the amplitude has the form
\bea
I_n & \equiv & -i\left({\mu\over 2}\right)^{4-D} 
\int {{\rm d}^D q\over (2\pi)^D}
\, 
{\bf q}^{2n} 
\({i\over {E\over 2} + q_0 -{ {\bf q}^2\over 2M} + i\epsilon}\)
\({i\over {E\over 2} - q_0 -{ {\bf q}^2\over 2M} + i\epsilon}\)
\nonumber\\
& = & -M (ME)^n (-ME-i\epsilon)^{(D-3)/2 } \ \ \Gamma\({3-D\over 2}\)
{\left({\mu\over 2}\right)^{4-D} \over  (4\pi)^{(D-1)/2}}
\ \ \ \  ,
\label{eq:loopi}
\eea
where $D$ is the number of space-time dimensions.
In the \pds scheme the pole at $D=3$ is removed by adding a local counterterm
to the lagrange density, so that the sum of the loop graph and counterterm 
in $D=4$ dimensions is
\be
I_n^{PDS} = I_n + \delta I_n = - (ME)^n \left({M\over 4\pi}\right) (\mu + ip).
\label{eq:ipds}
\ee

\begin{figure}[t]
\qquad\qquad\psfig{figure=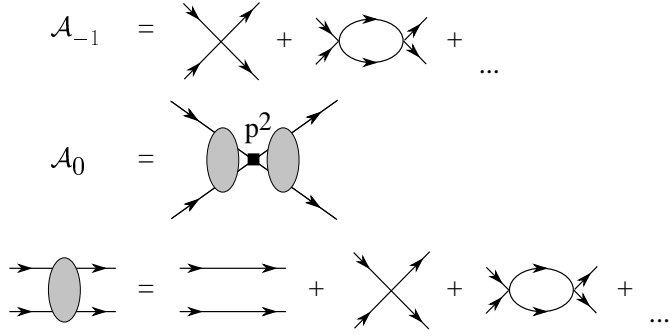,height=2.0in}
\caption{Leading and subleading contributions arising from local operators.
\label{KSW_fig2}}
\end{figure}

The amplitude $\CA$ is independent of the subtraction point $\mu$ and
this  determines the $\mu$ dependence of the coefficients, $C_{2n}$. 
In the \pds scheme one finds that for $\mu\gg 1/|a|$, 
the couplings $C_{2n}(\mu)$ scale as
\be
C_{2n}(\mu) \sim {4\pi \over M \Lambda^n \mu^{n+1}}\ ,
\label{eq:cscale}
\ee
so that if we take $\mu \sim p$, $C_{2n}(\mu)\sim 1/p^{n+1}$.  
A factor of $\nabla^{2n}$ at a vertex scales as $p^{2n}$, 
while each loop contributes a factor of  $p$.  
Therefore, the leading order contribution to the scattering amplitude $\CA_{-1}$  
scales as $p^{-1}$ and consists of the sum of bubble diagrams with $C_0$ vertices. 
Contributions  scaling as higher powers of $p$ come from perturbative 
insertions of derivative interactions, dressed to all orders by $C_0$.  
The first two  terms in the expansion
\bea
{\cal A}_{-1} & = &  { -C_0\over \left[1 + {C_0 M\over 4\pi} (\mu + ip)\right]}
\ \ \ ,\ \ \ 
{\cal A}_0  ={ -C_2 p^2\over \[1 + {C_0 M\over 4\pi}(\mu + ip)\]^2}
\ \ \ \ ,
\eea
correspond to the Feynman diagrams in Fig.~(\ref{KSW_fig2}).
A comparison with eq.(\ref{eq:aexp2}) gives
\bea
C_0(\mu) &=& {4\pi\over M }\({1\over -\mu+1/a}\)
\ \ \ , \ \ \ 
C_2(\mu) =  {4\pi\over M }\({1\over -\mu+1/a}\)^2 {{r}_0\over 2}
\ \ \ .
\label{eq:cvals}
\eea
The dependence of $C_{2n}(\mu)$ on
$\mu$ is determined by requiring  the amplitude be independent of the
renormalization scale  $\mu$.  
The physical parameters $a$, $ r_n$ enter as
boundary conditions on the resulting renormalization group (RG) 
equations. 
In general one finds the coefficients to be
\be
C_{2n}(\mu) = {4\pi \over M(-\mu+1/a)} \({r_0/2\over -\mu+1/a}\)^n + O(\mu^{-n})
\ \ \ ,
\ee
which has the scaling property in eq.(\ref{eq:cscale}).  
The leading behavior depends on the two parameters $a$ and $r_0$ 
encountered when solving for $C_0(\mu)$ and $C_2(\mu)$.  
This is due to the $C_{2n}$ couplings being driven primarily 
by lower dimensional interactions.

The inclusion of pions into the theory is straightforward.
While the coefficients of the local operators are renormalized,
and scale as powers of the renormalization scale $\mu$
(we use $Q \equiv \mu\sim p\sim m_q^{1/2}$),
the exchange of a single potential pion does not suffer from such
renormalizations and therefore pion exchange is
a sub-leading contribution, $Q^0$.
At the same order as the exchange of a potential pion is an
insertion of a $C_2$ operator and  a
single insertion of the quark mass matrix $m_q$. 
Ignoring isospin violation,  
these operators involving insertions of the 
light quark mass matrix with coefficients $D_2$
have the same structure as the $C_0$ operators.
For  $\mu\sim m_\pi$,
$C_0(\mu)\propto 1/\mu$, 
$C_2(\mu)\propto 1/\mu^2$ 
and $D_2(\mu)\propto  1/\mu^2$ 
for the $\si$ and $\siii$ channels.   
A  feature of the theory with pions is that this scaling 
behavior breaks down at low momentum, $p\sim 1/|a|$, 
and at sufficiently
high momentum.  
\begin{figure}[t]
\qquad\qquad\psfig{figure=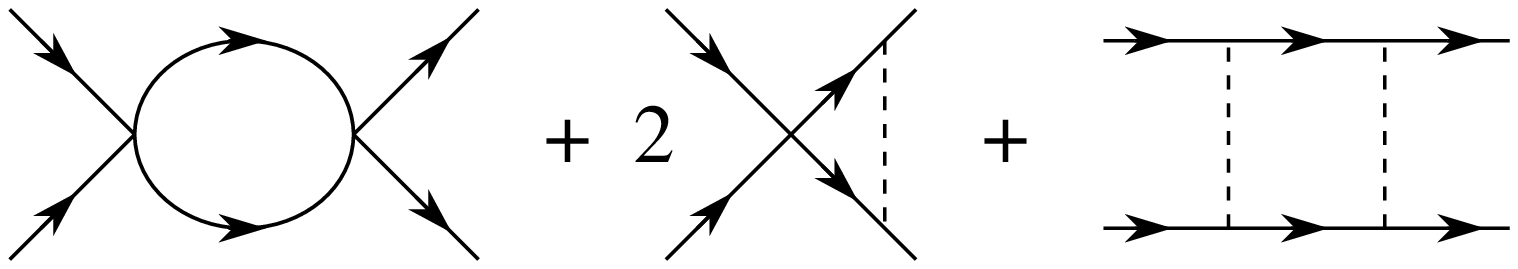,height=0.7in}
\caption{Contributions to the $\beta$-functions for $C_0$
in the theory with pions
\label{KSW_fig4}}
\end{figure}
Solving the RG equation arising from the graphs shown in Fig.~(\ref{KSW_fig4})
in the $\si$ channel with the boundary
condition $C_0^{(\si)}(0)=4\pi a_1/M$, 
where $a_1$ is the $\si$ scattering length,  we find  for 
$\mu\gg 1/|a_1|$
\be
C_0^{(\si)}(\mu)\simeq -{4\pi\over M\mu}\(1+{\mu\over \Lambda_{NN}}\)\ ,
\ee
with
\bea
\Lambda_{NN} & = & {8\pi f^2\over g_A^2 M} \sim 300\ {\rm MeV}
\ \ \ ,
\eea
and therefore the power counting changes when $\mu \sim \Lambda_{NN}$.
The 
UV fixed point toward which $C_0^{(\si)}$ is driven largely cancels 
the $\delta$-function component
of the single potential pion exchange in the $\si$ channel.  
As a result, this power counting is valid
up  to $p\sim \Lambda_{NN}$
and the power counting in both channels is  expected 
to fail at momenta on the order of $\Lambda_{NN}$.  
We conclude that the expansion parameter for this theory
is $\sim m_\pi/\Lambda_{NN}\sim {1\over 2}$, larger than one would like.

%%%%%%%%%%%%%%%%%%%%%%%%%%%%%%%%%%%%%%%%%%
\section{NN scattering in the $\si$ Channel}
\label{sec:5}
%%%%%%%%%%%%%%%%%%%%%%%%%%%%%%%%%%%%%%%%%%

\begin{figure}[t]
\qquad\psfig{figure=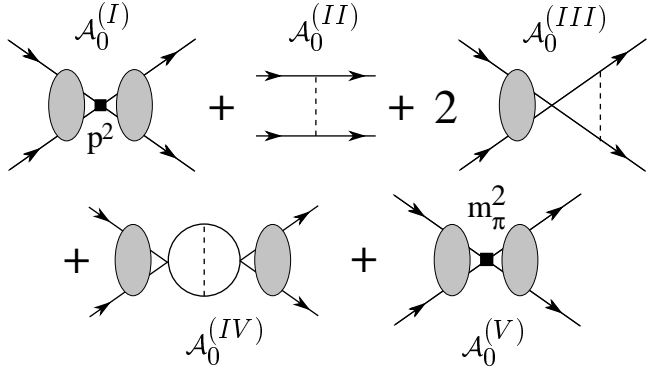,height=2.0in}
\caption{Graphs contributing to the subleading amplitude $\CA_{0}$.
The shaded ovals are defined in Fig.~(\ref{KSW_fig2}). 
\label{KSW_fig5}}
\end{figure}

Having established a consistent power-counting in the previous
sections we now apply it to $NN$ scattering in the $\si$ channel.
The amplitude at order $Q^{-1}$ and $Q^0$ determined 
from the graphs shown in
Fig.~(\ref{KSW_fig2}) and Fig.~(\ref{KSW_fig5})
are 
\bea
\CA_{-1} & = & 
-{ C^{(\si)}_0\over  1 + C^{(\si)}_0 {M\over 4\pi}  \left( \mu + i p \right) }
\ \ \ ,
\nonumber\\
 {\cal A}_0^{(I)} & = & 
-C_2^{(\si)} p^2
\left[ {\CA_{-1}\over C_0^{(\si)}  } \right]^2
\ \ \ ,
\nonumber\\
 {\cal A}_0^{(II)} &=&  \left({g_A^2\over 2f^2}\right) \left(-1 + {m_\pi^2\over
4p^2} \ln \left( 1 + {4p^2\over m_\pi^2}\right)\right)
\ \ \ \ ,
\nonumber\\
 {\cal A}_{0}^{(III)} &=& {g_A^2\over f^2} \left( {m_\pi M{\cal A}_{-1}\over 4\pi}
\right) \Bigg( - {(\mu + ip)\over m_\pi}
+ {m_\pi\over 2p} X(p,m_\pi)\Bigg)
\  \ \ \ ,
\nonumber\\
{\cal A}_0^{(IV)} &=& {g_A^2\over 2f^2} \left({m_\pi M{\cal A}_{-1}\over
4\pi}\right)^2 \Bigg(1 -\left({\mu + ip\over m_\pi}\right)^2
+ i X(p,m_\pi)  - \ln\left({m_\pi\over\mu}\right)  
\Bigg)
\ \ \ \ ,
\nonumber\\
{\cal A}_0^{(V)} &=& - D^{(\si)}_2 m_\pi^2 
\left[ {\CA_{-1}\over C_0^{(\si)}  }\right]^2
\ \ \ \  ,
\nonumber\\
X(p,m_\pi) & = & \tan^{-1} \left({2p\over m_\pi}\right) + {i\over 2} \ln
\left(1+ {4p^2\over m_\pi^2} \right)
\ \ \ \  .
\label{eq:amp1s0}
\eea
At order $Q^{-1}$ there is one unknown coefficient $C^{(\si)}_0$ that must be determined from 
data while at order $Q^0$ there are three unknown coefficients $C^{(\si)}_0, C^{(\si)}_2$ and
$D^{(\si)}_2$ that must be determined.
The graph giving $\CA_0^{(IV)}$ is divergent in four dimensions and therefore
gives rise to the logarithmic dependence on the renormalization scale $\mu$ 
in eq.~(\ref{eq:amp1s0}).
In order for the expansion to converge, 
the leading term $\CA_{-1}$ must capture most of the 
scattering length.
The phase shift $\delta$ is perturbatively expanded in $Q$,
$\delta = \delta^{(0)} + \delta^{(1)} + \ldots $.
\begin{figure}[t]
\qquad\qquad\qquad\psfig{figure=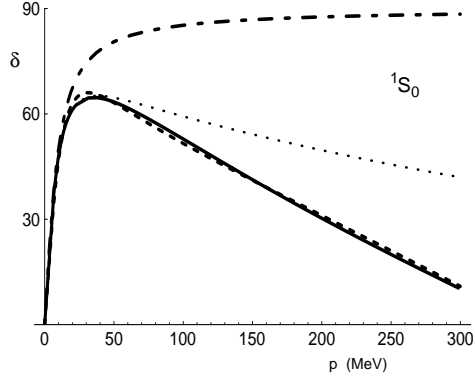,height=2.0in}
\caption{The phase shift $\delta$ for the $\si$ channel.
The dot-dashed curve is the  one parameter fit at order $Q^{-1}$,
that reproduces the scattering length.
The dashed curve corresponds to fitting $\delta$ 
between $0 < p < 200\ {\rm MeV}$, while the 
dotted curve corresponds to fitting the 
scattering length and effective range.
The solid line shows the results of the 
Nijmegen partial wave  analysis.
\label{KSW_fig6}}
\end{figure}
and fit to the results of the 
Nijmegan partial-wave analysis \cite{nijmegen}
over a momentum range $p\le 200 \MeV$. We find for $\mu=m_\pi$
\be
C_0^{(\si)} =-3.34\fm^2
\ ,\qquad
D_2^{(\si)} =-0.42\fm^4
\ ,\qquad
C_2^{(\si)} =3.24\fm^4
\ \ \  ,
\label{eq:numfit}
\ee
giving the dashed curve plotted in  Fig.~(\ref{KSW_fig6}).
It is clear from Fig.~(\ref{KSW_fig6}) that the corrections to the 
leading order result become substantial above $\sim 200\  {\rm MeV}$
and we expect the expansion to become unreliable at momenta larger
than this value.
We chose to renormalize at $\mu=m_\pi$ for our numerical analysis, but
we could have chosen 
any value of $\mu$, with $\Lambda_{NN}\gg\mu\gg 1/a$.  
The logarithm appearing in
the subleading amplitude suggests we choose $\mu\sim m_\pi$.

%%%%%%%%%%%%%%%%%%%%%%%%%%%%%%%%%%%%%%%%%%
\section{NN scattering in the $\siii-\diii$ Channel}
\label{sec:6}
%%%%%%%%%%%%%%%%%%%%%%%%%%%%%%%%%%%%%%%%%%

%
\begin{figure}[t]
\psfig{figure=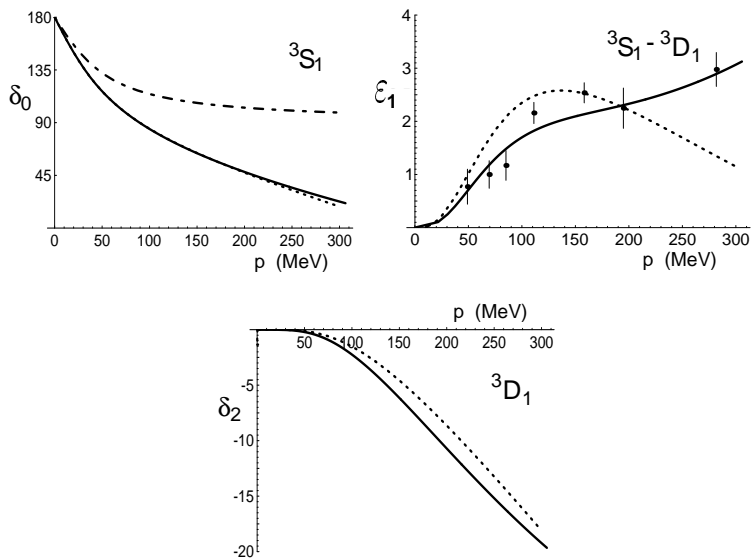,height=3.0in}
\caption{The phase shifts $\delta_0$, $\delta_2$ and  mixing parameter
$\varepsilon_1$ for the $\siii-\diii$ channel. 
The solid line denotes the results of the  Nijmegen partial wave  analysis.
The dot-dashed curve is the fit at order $Q^{-1}$ for $\delta_0$, while
$\delta_2 = \varepsilon_1 = 0$ at this order.
The dashed curves are the results of the order $Q^{0}$  fit of $\delta_0$ 
to the partial wave analysis over the momentum range $p\le 200\ \MeV$.
\label{KSW_fig7}}
\end{figure}

The analysis of scattering in the $\siii-\diii$ channel is a
straightforward extension of the analysis in the 
$\si$ channel.
The important difference is that the nucleons in the initial 
and final states with total angular momentum $J=1$ can be in an 
orbital angular momentum state of either $L=0$ or $L=2$.
The power counting for amplitudes that take the nucleons from a 
$\siii$-state to a $\siii$-state is identical to the analysis
in the $\si$-channel.
Operators  between two $\diii$ states are not directly renormalized by the 
leading operators, which project out only $\siii$ states.
However, they are renormalized by operators that mix the $\siii$ and $\diii$ 
states, which in turn are renormalized by the leading interactions.
Further, they involve a total of four spatial derivatives, two on the incoming 
nucleons, and  two on the out-going nucleons.   
Therefore, such operators contribute
at order $Q^3$, and can be neglected in the present computation.
Consequently, amplitudes for scattering from an $\diii$ state into an $\diii$ state
are dominated by single potential pion exchange at order $Q^0$.
Operators connecting $\diii$ and $\siii$ states
are renormalized by the leading  operators, but only on the 
$L=0$ ``side'' of the operator.
Therefore the coefficient of this operator, $C_2^{(\siii-\diii)}\sim 1/\mu$,
contributing at order $Q^1$ and  it can be neglected at order $Q^0$.
Thus,  mixing between $\diii$ and $\siii$ states is dominated by
single potential pion exchange dressed by a bubble chain of $C_0^{(\siii)}$ operators
and a parameter free prediction for this mixing 
exists at order $Q^0$.
Fitting the parameters $C_0^{(\siii)}$,  $C_2^{(\siii)}$
and  $D_2^{(\siii)}$  to the phase shift $\delta_0$
over the momentum range $p\le 200 \MeV$ yields, at $\mu=m_\pi$
\bea
C_0^{(\siii)} =-5.51\fm^2\ ,\ 
D_2^{(\siii)} =1.32\fm^4\ ,\ 
C_2^{(\siii)} =9.91\fm^4
\  .
\label{eq:numfitc}
\eea
The dashed curves in Fig.~(\ref{KSW_fig7})  show the  phase shifts $\delta_0$, $\delta_2$
and mixing parameter $\varepsilon_1 $
compared to the  Nijmegen partial wave  analysis \cite{nijmegen} for this set of
coefficients.
There are no free parameters at this order in either $\varepsilon_1 $ or $\delta_2 $
once $C_0^{(\siii)}$ has been determined from $\delta_0$.

%%%%%%%%%%%%%%%%%%%%%%%%%%%%%%%%%%%%%%%%%%
\section{The Deuteron}
\label{sec:7}
%%%%%%%%%%%%%%%%%%%%%%%%%%%%%%%%%%%%%%%%%%

%
\begin{figure}[t]
\psfig{figure=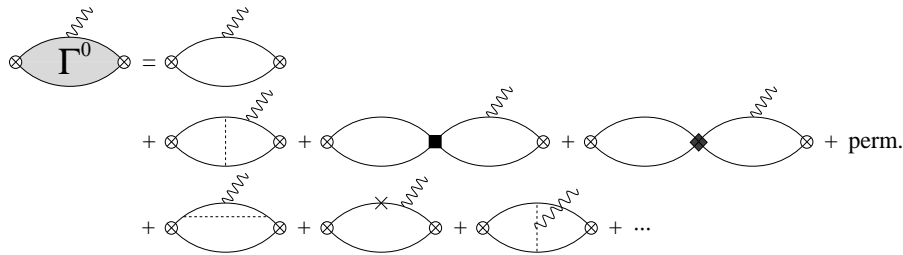,height=1.4in}
\caption{The diagrams contributing to the electric form factors of the deuteron.
\label{fig:gam0}}
\end{figure}
Once the Lagrange density in the nucleon sector has been
established the standard tools of field theory
can be used to determine the properties of
the deuteron\cite{KSWb}.
To compute the electromagnetic form factors of the deuteron one first computes
the three point correlation function between a source that creates a nucleon
pair in a $\siii$ state, a source that destroys a nucleon
pair in a $\siii$ state and a source that creates a photon.
After LSZ reduction and wavefunction renormalization one obtains the
electromagnetic form factors.
Leading, subleading and subsubleading order graphs contributing to the electric
form factors of the deuteron are shown in Fig.~(\ref{fig:gam0}).
\begin{figure}[t]
\qquad\qquad\psfig{figure=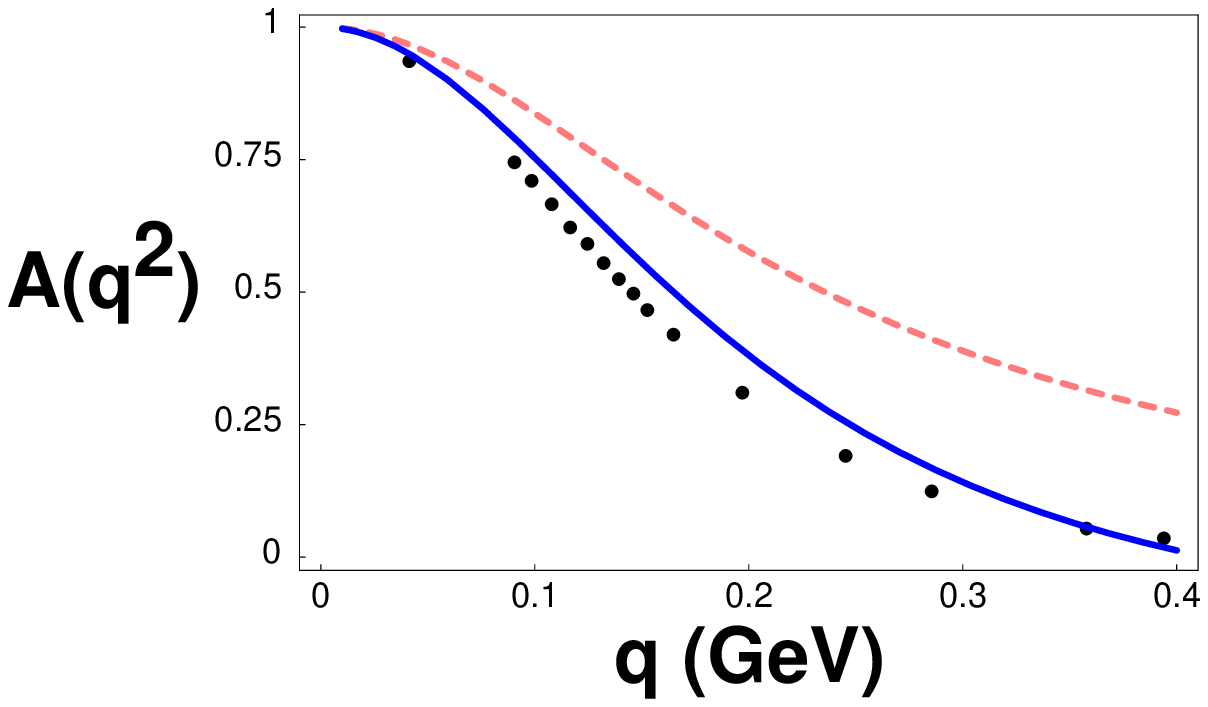,height=1.6in}

\qquad\qquad\psfig{figure=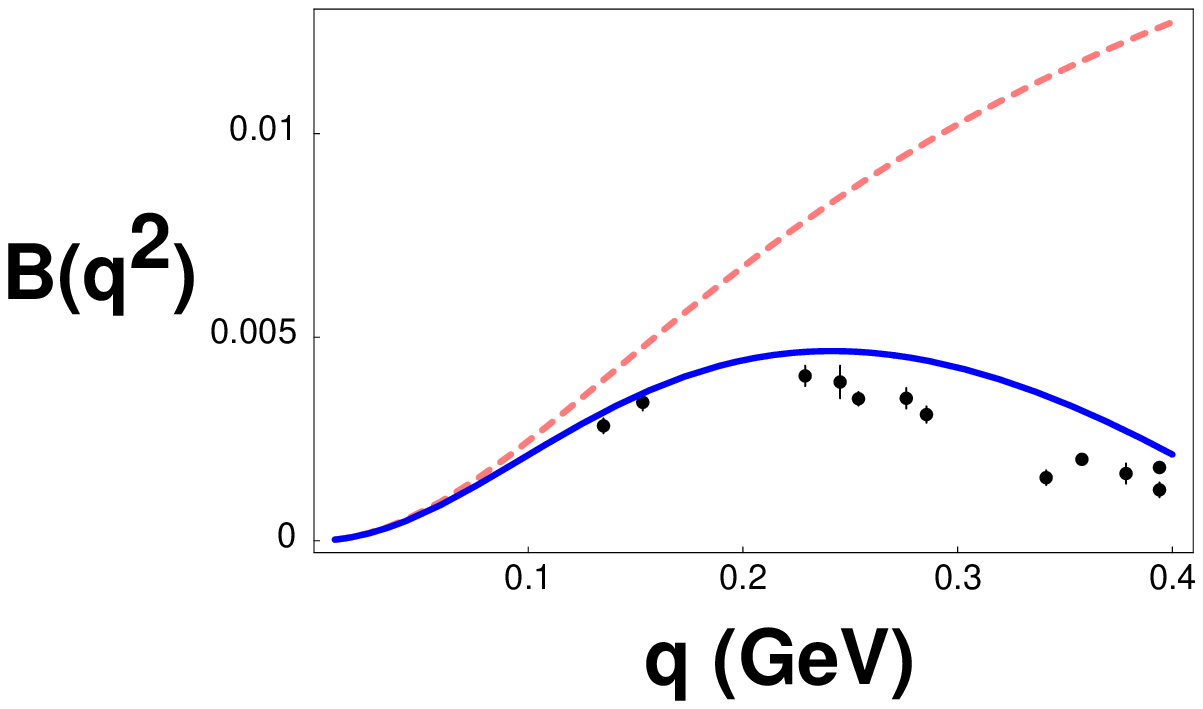,height=1.6in}
\caption{The form factors $A(q^2)$ and $B(q^2)$ measured in elastic
  electron-deuteron
  scattering.
  The dashed curve is the leading order prediction, while the solid
  curve subleading prediction.
  There is one counterterm in $B(q^2)$ that is fixed by the deuteron magnetic moment.
\label{fig:ABplot}}
\end{figure}
The resulting form factors $A(q^2)$ and $B(q^2)$ that appear in the differential
cross section for electron deuteron scattering are shown in
Fig.~(\ref{fig:ABplot}).
$A(q^2)$ is dominated by the charge form factor and $B(q^2)$ depends only upon
the magnetic form factor.
One sees that the form factors computed at subleading order agree well with the
data.
The charge radius of the deuteron is found to be
\begin{eqnarray}
  \sqrt{<r^2>}_{\rm th} & = & {1\over 2\sqrt{2}\gamma}
  \ +\
  C_2(\mu) {M (\mu-\gamma)^2\over 48\sqrt{2} \pi}
\ +\
{g_A^2 M m_\pi^2 (3 m_\pi + 10\gamma)\over 288\sqrt{2}\pi f_\pi^2
    (m_\pi + 2 \gamma)^3}\ + \ ...
  \nonumber\\
  & = &  \ 1.89\  {\rm fm} + ... 
  \nonumber\\
   \sqrt{<r^2>}_{\rm expt} & = & 1.963\  {\rm fm} 
  \ \ \  ,
\end{eqnarray}
where $\gamma=\sqrt{M_N B}$ is the binding momentum of the deuteron.
The result computed to subleading order agrees with the observed value.
The quadrupole moment vanishes at leading order in the expansion but receives
a contribution at subleading order from the exchange of one potential pion, giving
\begin{eqnarray}
  \mu_{{\cal Q}, th} & = & {g_A^2 M ( 6 M B + 9m_\pi\gamma + 4
    m_\pi^2)\over 30\pi f_\pi^2 (m_\pi + 2 \gamma)^3}
  \ + \ ....
  \nonumber\\
  & = & 0.40\ {\rm fm^2}\ + ... 
    \nonumber\\
  \mu_{{\cal Q}, expt} & = &  0.2859\ {\rm fm^2} 
\ \ \ ,
\end{eqnarray}
which is approximately $30\%$ larger than the experimental value.
Clearly, a subsubleading calculation is needed to ensure that the theoretical
value is converging to the experimental value.  This work is currently in
progress\cite{quadchen}.
In contrast, there is a local counterterm
contributing to the magnetic moment at subleading order. We find
\begin{eqnarray}
\mu_M & = & \mu_p\ +\ \mu_n\ +\ L_2 (\mu) {\gamma\over 2\pi} (\mu-\gamma)^2\ +\ ...
\nonumber\\
& = & 0.88\ -\  0.02\ \ \  (fit) 
\ \ \ \ ,
\end{eqnarray}
which determines the counterterm $L_2$ at the scale $\mu$.
Once this counterterm is determined from the deuteron magnetic moment, the
entire form factor $B(q^2)$ is determined to subleading order.

We have also computed the polarizabilities of the deuteron\cite{CGSS}, and
find that the scalar $\alpha_{E0}$ and tensor $\alpha_{E2}$ electric polarizabilities,
are
\begin{eqnarray}
  \alpha_{E0} & = & {\alpha M_N\over 32 \gamma^4}\ +\ 
{\alpha M_N^2\over 64\pi\gamma^3} 
  C_2(\mu) (\mu-\gamma)^2
  \nonumber\\
& + & {\alpha g_A^2 M_N^2\over 384\pi f^2}\ 
  { m_\pi^2 ( 3 m_\pi^2 + 16 m_\pi\gamma + 24\gamma^2)\over
      \gamma^3 (m_\pi+2 \gamma)^4}
    \ +\ ...
  \nonumber\\
\alpha_{E2} & = & - {\alpha g_A^2 M_N^2\over 80\pi f^2} \ 
  { 2m_\pi^3+11 m_\pi^2\gamma + 16 m_\pi\gamma^2 + 8\gamma^3\over \gamma^2
    (m_\pi+2\gamma)^4}
  \nonumber\\
 \alpha_{E0} & = & 0.595\ {\rm fm}^3\ +\ ...\ \ ,\ \ \alpha_{E2} =  -0.062\ {\rm fm}^3\ +\ ...
\ \ \ ,
\end{eqnarray}
while the scalar and tensor magnetic polarizabiltities, $\beta_{M0}$ and $\beta_{M2}$,
are
\begin{eqnarray}
  \beta_{M0}  & = & {\alpha\over 2 M_N}
  \left[ -{1\over 16\gamma^2} \ +\
    {2(\kappa^{(0)})^2 +(\kappa^{(1)})^2 \over 3\gamma^2}
    \ +\
    { (\kappa^{(1)})^2\over 6\pi}{M_N\over \gamma} {\cal A}_{-1}^{(\si)}(-B)
    \right]
  \nonumber\\
  \beta_{M2} & = & -{\alpha\over M_N}
  {(\kappa^{(0)})^2 -(\kappa^{(1)})^2 \over 2\gamma^2}
  \ +\
    { \alpha (\kappa^{(1)})^2\over 4\pi\gamma} {\cal A}_{-1}^{(\si)}(-B)  
  \nonumber\\
  \beta_{M0} & = & 0.067\ {\rm fm}^3\ +\ ...
  \ \ \ ,\ \ \ \beta_{M2} = 0.195\ {\rm fm}^3\ +\ ...
\ \ \ ,
\end{eqnarray}
where  ${\cal A}_{-1}^{(\si)}(E)  $ is the scattering amplitude in the $\si$
channel evaluated at a center of mass energy $E$.
$\kappa^{(0)}$ and $\kappa^{(1)}$ are the isoscalar and isovector magnetic
moments of the nucleon repectively.

%%%%%%%%%%   Limitations  %%%%%%%%%%%%%

\section{Present  Limitations}
\label{sec:8}

The power counting fails at the scale $\Lambda_{NN}$  and 
there has been little progress in understanding how to deal momentum higher than
$\Lambda_{NN}$ in the effective field theory framework.
Therefore we are presently unable to address some important areas,
such as photo-pion production off the deuteron or elastic pion deuteron
scattering.  The typical momentum scale in such a process is
$\sim\sqrt{M_N m_\pi}$ greater than $\Lambda_{NN}$.

It is also worth commenting on the role of   baryonic resonances in this program.
The impact of the $\Delta$ resonance has been determined in Weinberg's 
power-counting scheme
in two different prescriptions\cite{KoMany,Sa96}
\footnote{The $\Delta (1232)$ and other  baryon resonances have been 
consistently included in the single nucleon sector\cite{JMhung}.}.
It is found not to play an important role in $NN$ scattering as the  
mass scale that sets the size of its contribution is 
$\sqrt{M (M_\Delta-M)}\sim 500 \ {\rm MeV}$.
This scale is higher than the scale at which the power counting becomes inappropriate, 
$\Lambda_{NN}$, and so  the baryonic resonances should not be included, until the theory above
the scale $\Lambda_{NN}$ is constructed.

%%%%%%%%%%%%%%%%%%%%%%%%%%%%%%%%%%%%%%%%%%%%%%%%%%%%%%%%%%%
\section{Conclusions}
\label{sec:9}
%%%%%%%%%%%%%%%%%%%%%%%%%%%%%%%%%%%%%%%%%%%%%%%%%%%%%%%%%%%

After several years of investigation we
have constructed a consistent power counting for an effective field theory
description of the nucleon nucleon interaction\cite{KSWb}.
Pions are subleading compared to the local momentum independent
four-nucleon operators
and  can be treated in perturbation theory.
$NN$ scattering in the 
$\si$ and $\siii-\diii$ channels to sub-leading order has been considered and 
most impressive perhaps is the parameter-free prediction
of the $\siii-\diii$ mixing parameter $\varepsilon_1$
which agrees reasonably well with the Nijmegen phase shift analysis.
The properties of the deuteron bound state follow straightforwardly from this
construction.
I presented the electromagnetic form factors and
polarizabilities.
One of the nice features of having a field theory construction is that there
are many  well known theorems.  One such theorem\cite{Politzer,Arzt}
is that 
contributions from operators that vanish by the equations of motion are
of the form of higher dimension operators that do not vanish by the
equations of motion and therefore such operators can be neglected.

The future looks extremely promising for a systematic 
analysis of nuclear physics using effective field theory.
The short term program will be to continue to examine the two-body systems in
detail,
working to higher orders in the effective field theory expansion.
In the long-term one hopes to make progress in three-body\cite{Bvk}
and higher-body systems.
Both lines of investigation are
directly relevent to the BLAST program, and I look forward to close
cooperation between  experimental and theoretical endevours in this area.

\bigskip\bigskip

I would like to thank the organizers Ricardo Alarcon and
Richard Milner for putting together such a stimulating meeting.
This work is supported in part by 
Department of Energy Grant DE-FG03-97ER41014.

%%%%%%%%%%%%%%%%%%%%%%%%%%%%%%%%%%%%%%%%%%%%%%%%%%%%%%%%%%%

\end{document}